\documentclass[a4paper, oneside, twocolumn, notitlepage, 10pt]{extarticle_ecoc}
\usepackage{ecoc}
\usepackage{graphicx}
\usepackage{epstopdf} 

\usepackage[nolist,nohyperlinks]{acronym}

\begin{document}
\selectlanguage{american}    


\title{Deep Learning of Geometric Constellation Shaping including Fiber Nonlinearities}%


\author{
    Rasmus T. Jones\textsuperscript{(1)}, Tobias A. Eriksson\textsuperscript{(2)},
    Metodi P. Yankov\textsuperscript{(1,3)}, Darko Zibar\textsuperscript{(1)}
}

\maketitle                  


\begin{strip}
 \begin{author_descr}

   \textsuperscript{(1)} Department of Photonics Engineering, Technical University of Denmark,
   \textcolor{blue}{\uline{rajo@fotonik.dtu.dk}}

   \textsuperscript{(2)} Quantum ICT Advanced Development Center, National Institute of Information and Communications Technology (NICT), Tokyo, Japan
   
   \textsuperscript{(3)} Fingerprint Cards A/S, 2730 Herlev, Denmark

 \end{author_descr}
\end{strip}

\setstretch{1.1}


\begin{strip}
  \begin{ecoc_abstract}
A new geometric shaping method is proposed, leveraging unsupervised machine learning to optimize the constellation design. The learned constellation mitigates nonlinear effects with gains up to 0.13~bit/4D when trained with a simplified fiber channel model.
  \end{ecoc_abstract}
\end{strip}


\section{Introduction}
\begin{figure}[t]
   \centering
        \includegraphics[width=\linewidth]{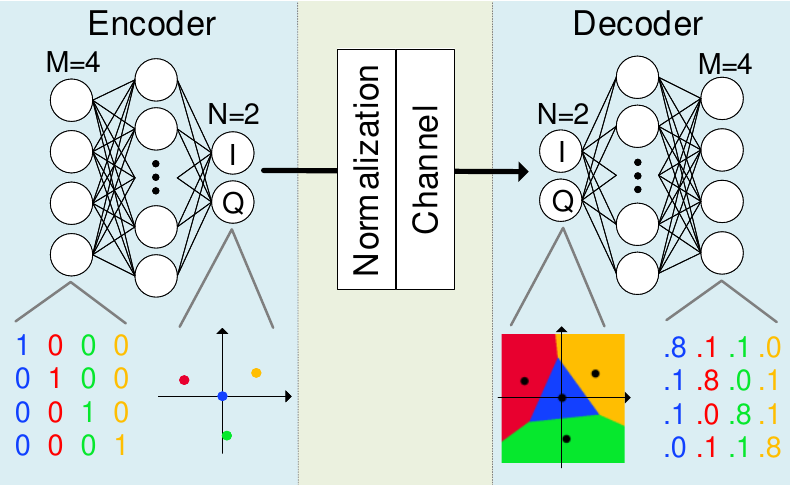}
    \caption{Trainable auto-encoder model.}
    \label{fig:autoEncoder}
\end{figure}

Optical transmission systems enabled modern data traffic applications, yet the future traffic growth is outpacing the achievable rates provided by such systems~\cite{ref1}.
Systems with high spectral efficiency are in demand, but limited by available \ac{SNR}.
For the \ac{AWGN} channel, constellation shaping, either geometric or probabilistic, provide up to 1.53 dB gain in \ac{SNR}~\cite{ref2}.
Both shaping methods have been applied in fiber optics but most often optimized under an \ac{AWGN} channel assumption~\cite{ref2,ref3}.
For the fiber optic channel, an optimal constellation is jointly robust to transmitter imperfections, amplification noise as well as signal dependent nonlinear effects~\cite{ref4}.
Especially, the modulation dependent nonlinear effects pose an intricate problem, since they are conditioned by the moment of the optimized constellation itself~\cite{ref10}. Constellation optimization for the nonlinear channel is thus non-trivial.
Machine learning is established for learning high dimensional relationships while taking various factors and constraints into account~\cite{ref5}.
O'Shea~et~al.~\cite{ref6} have shown the potential of learning codes by embedding a channel within an unsupervised machine learning algorithm, the auto-encoder~\cite{ref7}.
Similar, we propose, to embed fiber channel models within an auto-encoder~\cite{ref8} and learn geometric constellation shapes robust to the channel impairments, see Fig.~\ref{fig:autoEncoder}.
Thus, this method combines a channel model, the \ac{GN}-model~\cite{ref9} or the \ac{NLIN}-model~\cite{ref10}, with gradient based optimization from machine learning.
Together, a computational graph is composed with all the information inherit in the channel model and enough flexibility to optimize for a geometric constellation shape.
Trained on the \ac{GN}-model, the learned constellation is optimized for an \ac{AWGN} channel with an effective \ac{SNR}  governed by the launch power, thus nonlinear effects are not being mitigated. Trained on the \ac{NLIN}-model the learned constellation mitigates nonlinear effects by optimizing its moment.
The performance in \ac{MI} of the learned constellations is estimated with the \ac{NLIN}-model and simulations using the \ac{SSF} method.
Up to 0.13~bit/4D of gains are reported with respect to \ac{IPM}-based geometrically shaped constellations~\cite{ref3}.


\section{Channel Model and Auto-Encoder}
The unsupervised learning method embeds a channel model within an auto-encoder, as shown in Fig.~\ref{fig:autoEncoder}. The fiber channel is modeled by the \ac{NLIN}-model, where the channel impairments only depend on the \ac{ASE} noise, the average channel power $P$, and the 4th and 6th order moment ($\kappa$ and $\kappa_3$) of the constellation. The discrete \ac{NLIN}-model is described as follows~\cite{ref10}:
\begin{equation}
\begin{split}
y &= c(x,P^3,\kappa,\kappa_3)\\
&= x + n_{\text{ASE}} + n_{\text{NLIN}},
\end{split}
\end{equation}
where $y$ and $x$ are the received and transmitted symbols, $c(\cdot)$ the channel model, and $n_{\text{ASE}} \sim N(0,\sigma^2_{\text{ASE}})$ and $n_{\text{NLIN}} \sim N(0,\sigma^2_{\text{NLIN}}(P^3,\kappa,\kappa_3))$ are Gaussian noise samples with variance $\sigma^2_{\text{ASE}}$ and $\sigma^2_{\text{NLIN}}$, respectively. For the \ac{GN}-model the dependence on $\kappa$ and $\kappa_3$ is dropped, removing the modulation dependent nonlinear effect.
Wrapping the channel model with encoder and decoder, both as \ac{NN}, constructs the trainable auto-encoder model. \ac{NN}s operate with vectors of real numbers, thus the symbols $x$ and $y$ are transcribed into vectors of their real and imaginary part. The auto-encoder model is defined as follows:
\begin{equation}
\begin{split}
    \vec{x} &= f(\vec{s}),\\
    \vec{y} &= c(\vec{x},P^3,\kappa,\kappa_3),\\
    \vec{r} &= g(\vec{y}),\\
    L&(\vec{s},\vec{r}=g(c(f(\vec{s})))),
    \end{split}
\end{equation}
where $f(\cdot)$ is the encoder \ac{NN}, $g(\cdot)$ the decoder \ac{NN}. The goal is to reproduce the input $\vec{s}$ at the output $\vec{r}$ through a latent variable $\vec{x}$ (and its impaired version $\vec{y}$). This is achieved by minimizing a loss function $L(\cdot)$ such that $\vec{s} \approx \vec{r}$. 
An $N$ dimensional constellation is obtained with $\vec{x}$ chosen as $N$ dimensional vector. A constellation of order $M$ is enforced by training the auto-encoder with one-hot encoded vectors $\vec{s} \in S=\{ \vec{e}_i~|~i=1..M\}$, where $\vec{e}_i$ is the all zero vector except at row $i$, and which implies $|S|=M$.
The model is trained identical to a traditional auto-encoder. Multiple instances of $\vec{s}$ are uniformly sampled from $S$ and propagated through the auto-encoder model. The error obtained through the loss function is then backpropagated to optimize the \ac{NN} weights. The optimization is gradient based and step wise, which means, the free parameters of the encoder and decoder are optimized iteratively towards a state where the input is reproduced at the output. 
For further illustration $N$=2 and $M$=4 are chosen, see Fig.~\ref{fig:autoEncoder}. Thus, an instance of $\vec{x}$ represents a point as \ac{IQ} components and $\vec{s}$ is uniformly sampled from one-hot encoded vectors of length 4. In that sense, $\vec{s}$ represents the source of the system, but does not determine the actual constellation.
The encoder learns a bijective projection from $\vec{s}$ to $\vec{x}$, thus for every element in $S$, one point in the IQ plane is obtained. All points together resembe a constellation of order 4.
After the channel, the decoder classifies the impaired symbols $\vec{y}$ back to estimates of one-hot encoded vectors $\vec{r}$. The decoder has learned decision boundaries in between the impaired symbols, at the same time as the encoder was forced to produce a distinguishable set of symbols given the channel impairment.

\section{Numerical Simulation}
The auto-encoder method described above is used to find the optimal constellation for both, a channel governed by the \ac{NLIN}-model and the \ac{GN}-model. The auto-encoder optimizes the constellation but thereafter only the learned constellation is used, meaning an \ac{NN} is neither required at the deployed transmitter nor receiver.
The auto-encoder model is trained for each tested transmission distance and launch power.
The performance of the learned constellations is validated using both the \ac{NLIN}-model and \ac{SSF} simulations. The \ac{SSF} method simulates a dual polarization WDM system of 5 channels with 50~GHz channel spacing, each channel is root raised cosine shaped with 0.05~roll-off factor and 32~GHz bandwidth. The propagation is governed by 0.2~dB/km attenuation, dispersion coefficient 16.46~ps/(nm~km), nonlinear coefficient 1.3~(W~km)$^{-1}$ , and 20 spans of 100~km length. The power level is swept from -5~dBm to 5~dBm. The \ac{NLIN}-model, which assumes Nyquist shaped pulses, is used for all other performance estimations with 5 to 55 number of spans.
The results are compared to the performance of $M$-QAM constellations and \ac{IPM} type geometrically shaped constellations~\cite{ref3}.

\begin{figure*}[t]
    \centering
    \begin{minipage}{.5\textwidth}
        \centering
        \includegraphics[width=\linewidth]{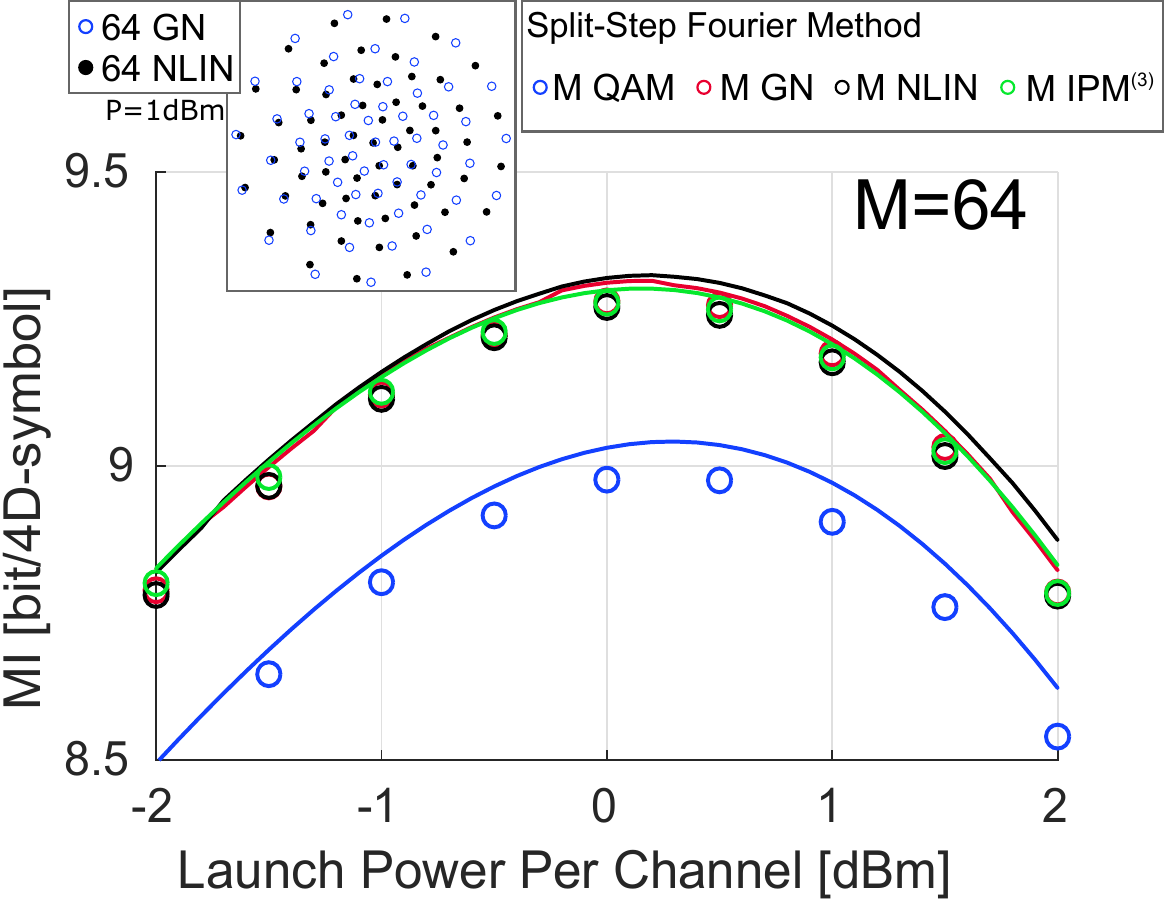}
    \end{minipage}%
    \begin{minipage}{0.5\textwidth}
        \centering
        \includegraphics[width=\linewidth]{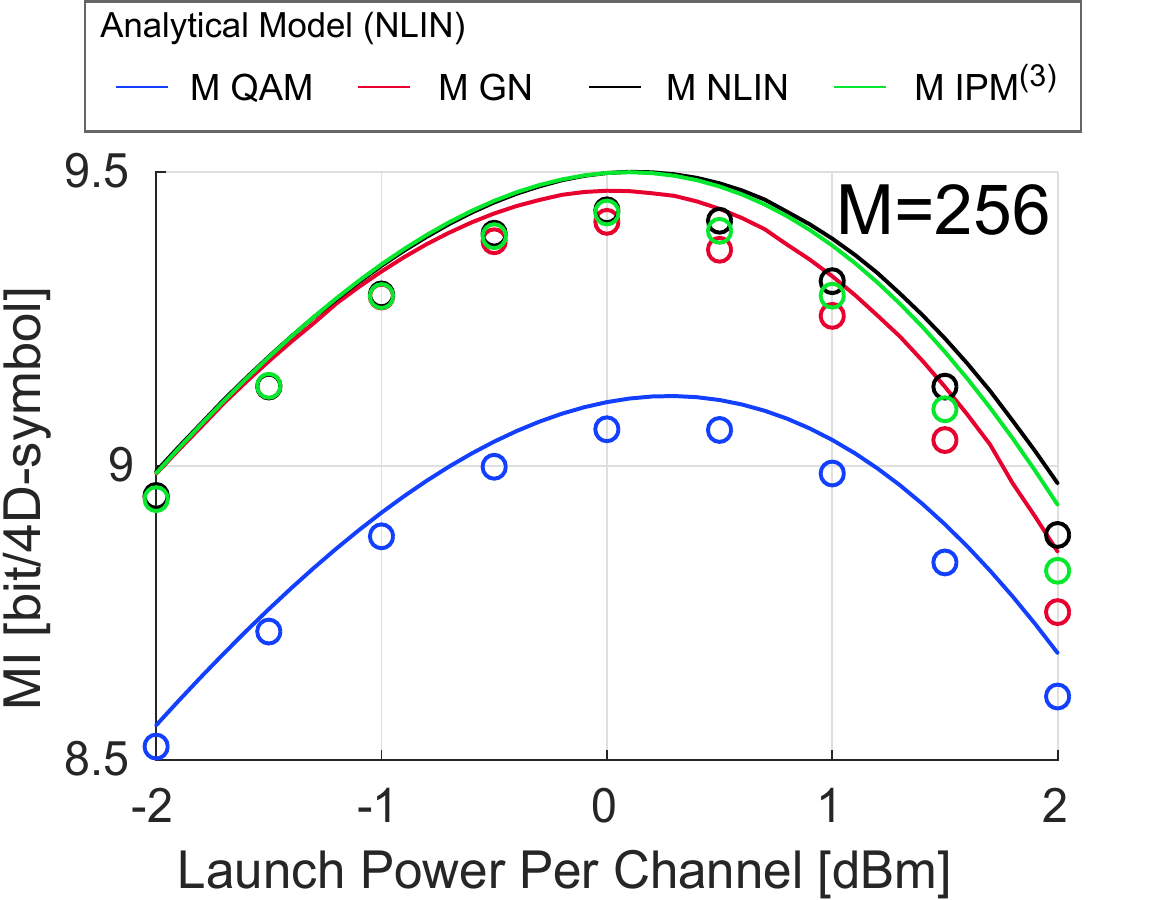}
    \end{minipage}
    \caption{Performance in \ac{MI} with respect to launch power after 2000~km transmission (20 spans) for $M$=64~\textbf{(left)} and $M$=256~\textbf{(right)}. Plots denoted as "$M$~\ac{GN}" and "$M$~\ac{NLIN}" indicate that the constellation was learned using the \ac{GN}-model and \ac{NLIN}-model, respectively. Lines depict performance evaluations using the \ac{NLIN}-model and markers using the \ac{SSF} method. (\textbf{top-left}) Learned $M$=64 constellation for 2000~km transmission and 1 dBm launch power.}
    \label{fig:allPlots}
\end{figure*}

\section{Results and Discussion}
\begin{figure}[t]
   \centering
        \includegraphics[width=\linewidth]{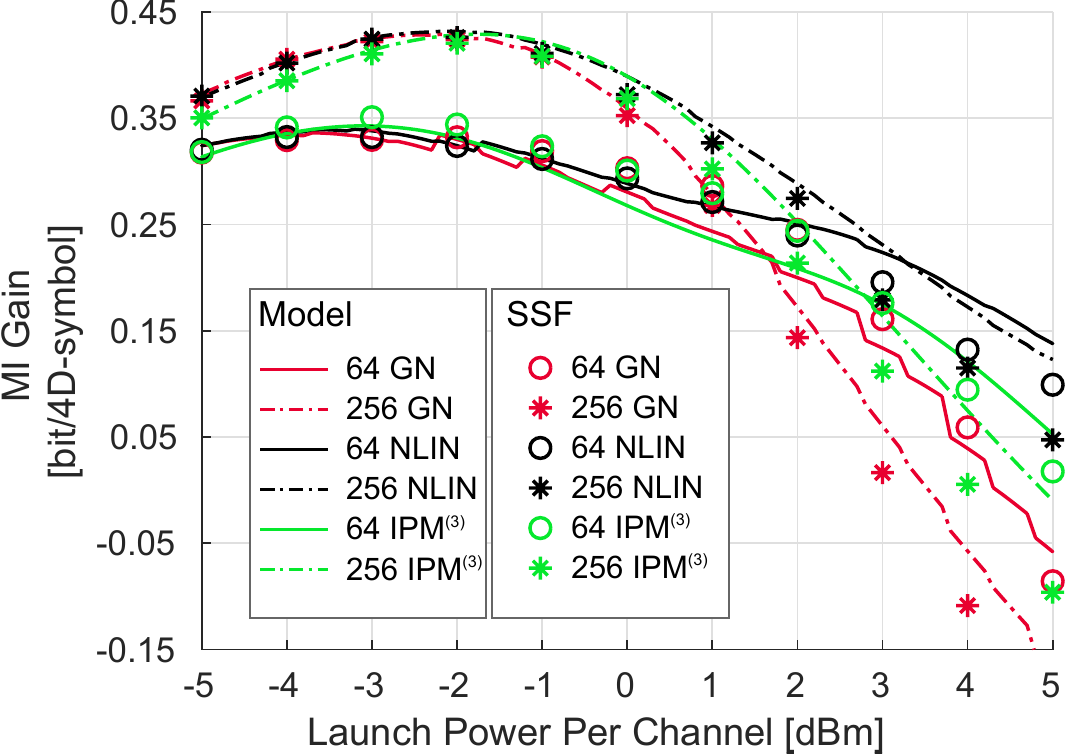}
    \caption{Gain compared to the standard $M$-QAM constellations with respect to launch power after 2000~km transmission (20 spans).}
    \label{fig:deltaMI}
\end{figure}

\begin{figure}[t]
   \centering
        \includegraphics[width=\linewidth]{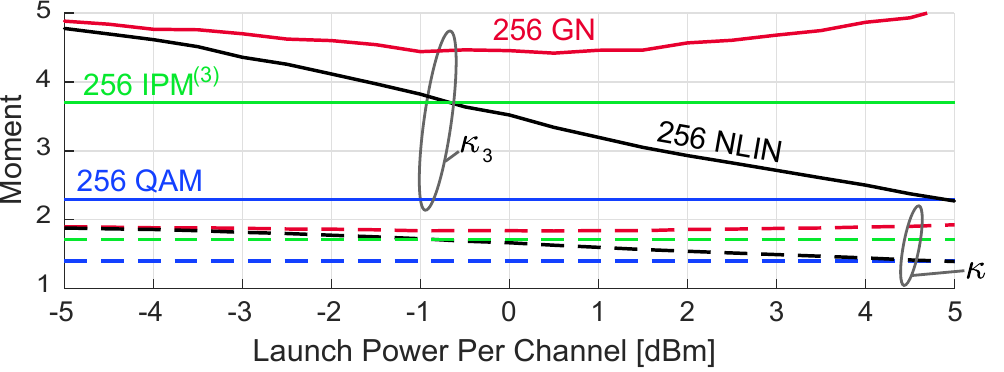}
        \includegraphics[width=\linewidth]{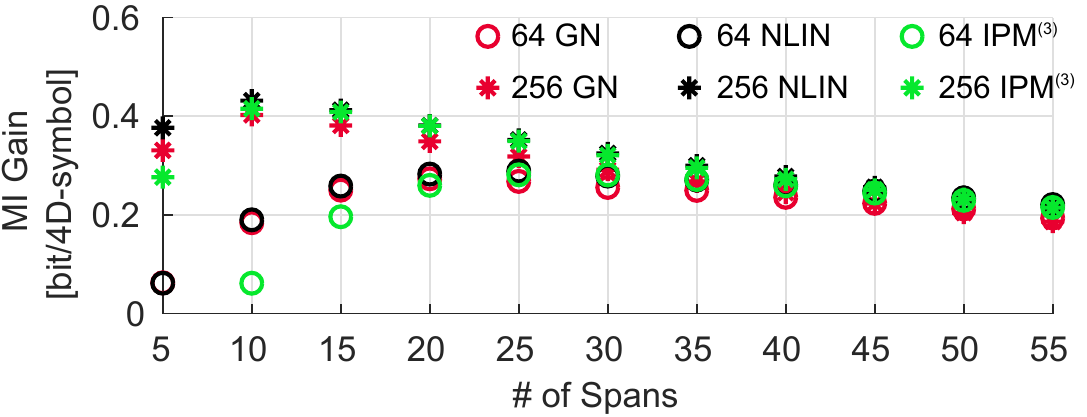}
    \caption{\textbf{(top)} 4th and 6th order moment ($\kappa$ and $\kappa_3$) of constellations with $M$=256 with respect to launch power optimized for 2000~km transmission distance.\\
    \textbf{(bottom)} Gain compared to the standard $M$-QAM constellations at the respective optimal launch power with respect to number of spans.}
    \label{fig:moment}
\end{figure}

The simulation results for a 2000~km transmission (20~spans) are shown in Fig.~\ref{fig:allPlots}.
The auto-encoder constellations obtain improved performance in \ac{MI} compared to standard QAM constellations.
Further, the optimal launch power for constellations trained on the \ac{NLIN}-model is shifted towards higher powers compared to the \ac{GN}-model and \ac{IPM}-based constellations.
In Fig.~\ref{fig:deltaMI}, at higher power the gain with respect to standard QAM constellations is larger for the \ac{NLIN}-model obtained constellations, and in Fig.~\ref{fig:moment}~(top), the constellations learned with the \ac{NLIN}-model have a negative slope in moment.
This shows that the auto-encoder wrapping the \ac{NLIN}-model, learns constellations causing less nonlinearities by reducing the moments.
At optimal launch power and transmission distances of 2500~km to 5500~km, the difference between \ac{NLIN}-model, \ac{GN}-model and \ac{IPM}-based constellations are marginal, Fig.~\ref{fig:moment} (bottom).
However, at lower transmission distance the auto-encoder learned constellations outperform the \ac{IPM}-based constellations by up to 0.13 bit/4D, since with less accumulated dispersion the signal dependent nonlinearities prevail~\cite{ref10}.
The marginal improvement over state of the art constellations at longer transmission distance suggests that temporal effects must be taken into account for larger gains.

\section{Conclusions}
An optimization method for geometric constellation shapes is proposed, by the means of an unsupervised learning method known as auto-encoder. With a channel model including modulation dependent nonlinear effects, the deep learning algorithm yields a constellation mitigating these, with gains up to 0.13~bit/4D. The method is used as is without any further analytic derivations necessary, since the machine learning optimization method is agnostic to the embedded channel model.

\begin{spacing}{0.7}
\vspace{-3pt}
\section{Acknowledgements}
\vspace{6pt}
\begin{footnotesize}
    We thank Dr. Tobias Fehenberger for discussions on the used channel models. This work was financially supported by Keysight Technologies (Germany, B\"oblingen).
\end{footnotesize}
\end{spacing}
\vspace{6pt}

\setlength{\bibsep}{3pt}
\bibliographystyle{abbrv}
\begin{spacing}{0.7}

\end{spacing}
\vspace{-4mm}

\acrodef{AWGN}[AWGN]{additive white Gaussian noise}
\acrodef{SNR}[SNR]{signal-to-noise ratio}
\acrodef{GN}[GN]{Gaussian noise}
\acrodef{NLIN}[NLIN]{nonlinear interference noise}
\acrodef{MI}[MI]{mutual information}
\acrodef{SSF}[SSF]{split-step Fourier}
\acrodef{ASE}[ASE]{amplified spontaneous emission}
\acrodef{IQ}[IQ]{in-phase and quadrature}
\acrodef{WDM}[WDM]{wavelength-division multiplexing}
\acrodef{QAM}[QAM]{quadrature amplitude modulation}
\acrodef{NN}[NN]{neural networks}
\acrodef{IPM}[IPM]{iterative polar modulation}
\end{document}